\documentclass[pra,aps,twocolumn,superscriptaddress]{revtex4}
\usepackage{graphics}

\begin{document}

\title{Experimental demonstration of teleportation-based programmable quantum gate}

\author{Luk\'a\v{s} Slodi\v{c}ka}
\affiliation{Department of Optics, Palack\'{y} University,
17. listopadu 50, 77200 Olomouc, Czech Republic}
\affiliation{Institute for Experimental Physics, University of Innsbruck, Technikerstr.
25, A-6020 Innsbruck, Austria}

\author{Miroslav Je\v{z}ek}
% \email{jezek@optics.upol.cz}
\affiliation{Department of Optics, Palack\'{y} University,
17. listopadu 50, 77200 Olomouc, Czech Republic}

\author{Jarom\'{i}r Fiur\'{a}\v{s}ek}
\affiliation{Department of Optics, Palack\'{y} University,
17. listopadu 50, 77200 Olomouc, Czech Republic}

\date{\today}

\begin{abstract}
We experimentally demonstrate a programmable quantum gate that applies a
sign flip operation to data qubit in an arbitrary basis fully specified
by the quantum state of a two-qubit program register. Our linear-optical
implementation is inspired by teleportation-based schemes for quantum
computing and relies on multiphoton interference and coincidence
detection. We confirm the programmability of the gate
and its high-fidelity performance by carrying out full quantum process
tomography of the resulting operation on data qubit for several
different programs.
\end{abstract}


\maketitle

Quantum computers can handle some computational tasks---such as
factoring large numbers \cite{Shor1994} and searching unsorted data
\cite{Grover1997}---more efficiently than classical computers can
do. Several feasible paradigms of quantum computation have been
proposed \cite{Gottesman99,KLM01,Raussendorf01,qcomputing} and selected
elementary quantum logic gates and processors have been demonstrated
\cite{gateexperiments}. Depending on what computational task is
desired, a particular dedicated processor has to be used.
By contrast, classical computers offer much more flexibility
for a fixed hardware, where the operation to be performed
on bits of a data register is determined by a suitable program
encoded in a program register.

In a seminal paper, Nielsen and Chuang \cite{Nielsen97} extended
the concept of programmable processors to quantum computation.
They considered a universal quantum gate that can be programmed
to perform an arbitrary unitary operation on the data qubits.
It turns out that, assuming a finite size quantum program register,
such universal gate can be implemented without errors only in
a probabilistic fashion \cite{Nielsen97,Hillery06}. 
Despite their probabilistic nature, programmable gates are
remarkable since a complete information on the implemented
quantum operation, which requires infinitely many classical bits
to specify, can be encoded into a finite number of quantum bits. 
Recently, a proof-of-principle experimental realization of a
probabilistic programmable quantum gate was demonstrated
\cite{Micuda08}, where a single program qubit determines a phase
shift applied between fixed basis states of a data qubit
\cite{Vidal02}. As a natural further step we can consider
a programmable gate where the computational basis itself is
also specified by the quantum state of the program register.
This is equivalent to programming a full single-qubit unitary operation,
which requires at least two-qubit program \cite{Nielsen97}.

\begin{figure}[!b!]
\hspace*{2mm}\centerline{\scalebox{0.85}{\includegraphics{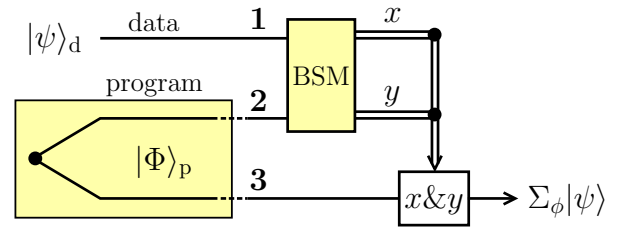}}}
\caption{Quantum circuit for teleportation-based programmable
quantum logic gate. Time proceeds from left to right. The single wires carry
quantum bits (qubits) and the double wires carry classical bits. Qubits
$2$ and $3$ form a program register prepared in a particular entangled
state $|\Phi\rangle_{\rm p}$. Bell-state measurement (BSM) is applied
to data qubit $1$ in an unknown quantum state $|\psi\rangle_{\rm d}$
and qubit $2$ of the program register. For singlet Bell state detected
in channels $1$ and $2$ the measurement yields $x=y=1$ which heralds
the successful gate operation leaving the remaining qubit $3$ in the
target state $\Sigma_\phi|\psi\rangle$. The unitary operation
$\Sigma_\phi$ is determined by the state of the program register.}
\label{sigmagate_fig1}
\end{figure}

Here, we report on a linear-optics experimental realization of
a high fidelity programmable quantum gate with a two-qubit program register.
The gate can perform a sign flip operation (SIGN) on the data qubit in an
arbitrary computational basis $\{|\phi\rangle,|\phi_\perp\rangle\}$ \cite{Hillery02},  
\begin{equation} \label{SIGN}
  \Sigma_{\phi} |\phi\rangle = |\phi\rangle, \,\quad
  \Sigma_{\phi} |\phi_{\perp}\rangle = - |\phi_{\perp}\rangle,
\end{equation}
where $\langle \phi|\phi_\perp\rangle=0$.
The computational basis in which the gate performs the sign flip is fully
specified by the quantum state of the program register. In order to encode
a particular basis $\{|\phi\rangle,|\phi_\perp\rangle\}$, the program
qubits should be prepared in a symmetric maximally entangled state 
\begin{equation} \label{program_register}
  |\Phi\rangle_{\rm p} =
  \frac{1}{\sqrt{2}}(|\phi\phi_{\perp}\rangle+|\phi_{\perp}\phi\rangle).
\end{equation}
We utilize the scheme of universal quantum computing of Gottesman and Chuang
\cite{Gottesman99} based on generalized quantum teleportation \cite{Bennett93}
that was recently used to demonstrate teleportation-based controlled-NOT gate
\cite{Pan08}. We apply Bell-state measurement (BSM) on the data qubit
$|\psi\rangle_{\rm d}=\alpha|\phi\rangle+\beta|\phi_\perp\rangle$ 
and one qubit from the program register, see Fig.~\ref{sigmagate_fig1}.
Particularly, if the qubits $1$ and $2$ are projected onto the maximally
entangled singlet state  
\begin{equation} \label{siglet_state}
|\Psi^{-}\rangle = \frac{1}{\sqrt{2}}(|01\rangle-|10\rangle)\equiv 
\frac{1}{\sqrt{2}}(|\phi\phi_\perp \rangle-|\phi_\perp\phi\rangle),
\end{equation}
then the remaining qubit $3$ is prepared in the sign-flipped state
$|\psi_{\rm out}\rangle_{\rm d}=\alpha|\phi\rangle-\beta|\phi_\perp\rangle$.
The success probability of the gate is thus $1/4$. The success
probability of the scheme for programmable SIGN gate proposed in
Ref.~\cite{Hillery02} is $1/3$. However, this latter approach requires
more complex state of program register and four controlled-NOT gates
between data and program qubits. Our implementation is significantly
less demanding at a cost of only slightly reduced success rate.

\begin{figure}[!bt!]
%\vspace*{-3mm}
\centerline{\scalebox{0.4}{\includegraphics{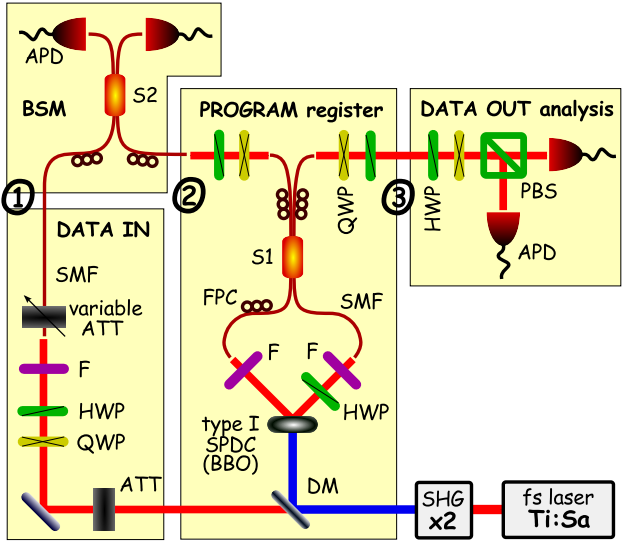}}}
\caption{A scheme of the experimental setup.
The initial subpicosecond pulses generated by titan-sapphire laser (Ti:Sa) 
at the fundamental wavelength of $800\,$nm are frequency doubled to $400\,$nm
by the process of second harmonic generation (SHG) and separated
from the fundamental ones by dichroic mirror (DM). 
The frequency doubled pulses pump a $2\,$mm thick nonlinear crystal
of $\beta$-barium borate (BBO) cut for non-collinear type-I spontaneous
parametric down-conversion (SPDC). The down-converted correlated photons
$2$ and $3$ at 800\,nm are filtered by narrow-band 1.75\,nm FWHM interference
filters (F), coupled to single-mode optical fibers (SMF), and mixed
by 50/50 fiber beam splitter (S1) which assures the perfect mode matching
of the photon modes. Polarization states of the photons are set to be
orthogonal before the splitter. The occurrence of one photon in each
output port of the splitter yields the entangled polarization
singlet state of the program register. It is further set by half-wave
(HWP) and quarter-wave plates (QWP) to any chosen entangled state.
The data qubit $1$ is encoded into a polarization state of a single
photon from the attenuated (ATT) fundamental laser pulse, frequency
filtered, and coupled to single-mode fiber. The data photon interferes
with photon $2$ from the program register at the second 50/50 fiber
beam splitter (S2) and the coincidence events are detected at the
output by single-photon avalanche photodiodes (APD) and multichannel
coincidence logic. A polarization state of the remaining
photon $3$ carrying output data qubit is simultaneously analyzed
using wave plates and polarizing beam splitter (PBS). 
Fixed unitary polarization transformations induced by optical fibers
are compensated by fiber polarization controllers (FPC). The photon
arrival times are precisely synchronized by optical delays not shown
in the scheme.}
\label{sigmagate_fig2}
\end{figure}

The experimental scheme relies on double Hong-Ou-Mandel interference
\cite{HOM} of three photons generated by separated sources of different
photo-statistics \cite{threephotons}. The input data qubit $1$
is encoded into a polarization state of a single photon from an
attenuated coherent laser pulse whereas program-register qubits
$2$ and $3$ are carried by polarization states of signal and idler
photons prepared by pulsed spontaneous parametric down-conversion,
see Fig.~\ref{sigmagate_fig2}. The down-converted pair of photons with
perpendicular linear polarizations set by wave-plates is coupled into
single-mode fibers and interferes at a balanced fiber beam splitter
S1 producing the maximally entangled polarization singlet state
$|\Psi^{-}\rangle_{23}$, that is transformed to any desired
program-register state (\ref{program_register}) easily by local
unitary operations. We independently verified the correctness of the
program-state preparation by means of full quantum state tomography
\cite{statetomo}. Photons $1$ and $2$ then interfere on the second
balanced fiber splitter S2 and they are detected in coincidence basis.
The joint detection event represents partial BSM that projects the
photons $1$ and $2$ onto the singlet state $|\Psi^{-}\rangle$.

The desired SIGN operation of the gate in particular basis is
analyzed by input-output probing for several chosen program-register
states. The data photon is prepared in one of the four input states 
$|H\rangle$ (linear horizontally polarized), 
$|V\rangle$ (linear vertically polarized),
$|D\rangle = \frac{1}{\sqrt{2}}(|H\rangle + |V\rangle)$
(linear diagonally polarized), and
$|R\rangle = \frac{1}{\sqrt{2}}(|H\rangle + {\rm i}|V\rangle)$
(right-circularly polarized). The state of the photon $3$ at the
output of the gate is measured in three mutually unbiased
polarization bases $\{|H\rangle$, $|V\rangle\}$,
$\{|D\rangle$, $|A\rangle\}$, and $\{|R\rangle$, $|L\rangle\}$,
where $|A\rangle=\frac{1}{\sqrt{2}}(|H\rangle - |V\rangle)$ and
$|L\rangle = \frac{1}{\sqrt{2}}(|H\rangle - {\rm i}|V\rangle)$.
This yields altogether $24$ characteristic threefold coincidence
rates. The raw data were corrected for the different
detection efficiencies of the employed detectors and for accidental
counts, which have been both measured prior to the main experiment.

From the experimental data we reconstruct the quantum operation that
fully characterizes the transformation of the data qubit for a fixed
state of the program register. For this purpose we exploit the
Jamiolkowski-Choi isomorphism \cite{Jamiolkowski72} which tells us
that every single-qubit quantum operation $\mathcal{E}$ can be represented
by a positive semidefinite operator $\chi$ on Hilbert space of two qubits.
This operator can be given a clear and intuitive physical meaning, $\chi$
is a density matrix of a two-qubit state obtained from the initial maximally
entangled state $|\Phi^{+}\rangle=|HH\rangle+|VV\rangle$ by applying the
operation $\mathcal{E}$ to one of the qubits. We have
$\chi=\mathcal{I}\otimes\mathcal{E}(|\Phi^{+}\rangle\langle\Phi^{+}|)$, where
$\mathcal{I}$ denotes the identity operation. Thus $\chi$ is a square positive
semidefinite Hermitean matrix with four columns and rows. The condition
$\chi\geq 0$ is equivalent to the fact that the operation is physical,
i.e. that the map $\mathcal{E}$ is completely positive.
Given a density matrix of the input state $\rho_{\mathrm{in}}$, the output
density matrix can be calculated according to the formula 
$\rho_{\mathrm{out}}=\mathrm{Tr}_{\mathrm{in}}
[(\rho_{\mathrm{in}}^{T}\otimes \openone_{\mathrm{out}})\,\chi]$,
where $T$ denotes transposition in a fixed basis, e.g. $\{|H\rangle,|V\rangle\}$. 
For each input state $\rho_{j}$ the probability of a particular measurement
outcome associated with projector $\Pi_{k}$ on the output state is given by
$p_{jk}=\mathrm{Tr}[(\rho_{j}^T\otimes \Pi_k)\,\chi]$. From the 
measured coincidence rates that approximate the theoretical probabilities
$p_{jk}$ we determine the quantum process matrix $\chi$ by an iterative
maximum-likelihood estimation algorithm that is described in detail
elsewhere \cite{processtomo}. This statistical reconstruction method yields
a quantum process matrix $\chi$ that is most likely to produce the observed
experimental data and, simultaneously, assures complete positivity
of the reconstructed operation.

\begin{figure}[!tbh!]
\centerline{\scalebox{1.1}{\includegraphics{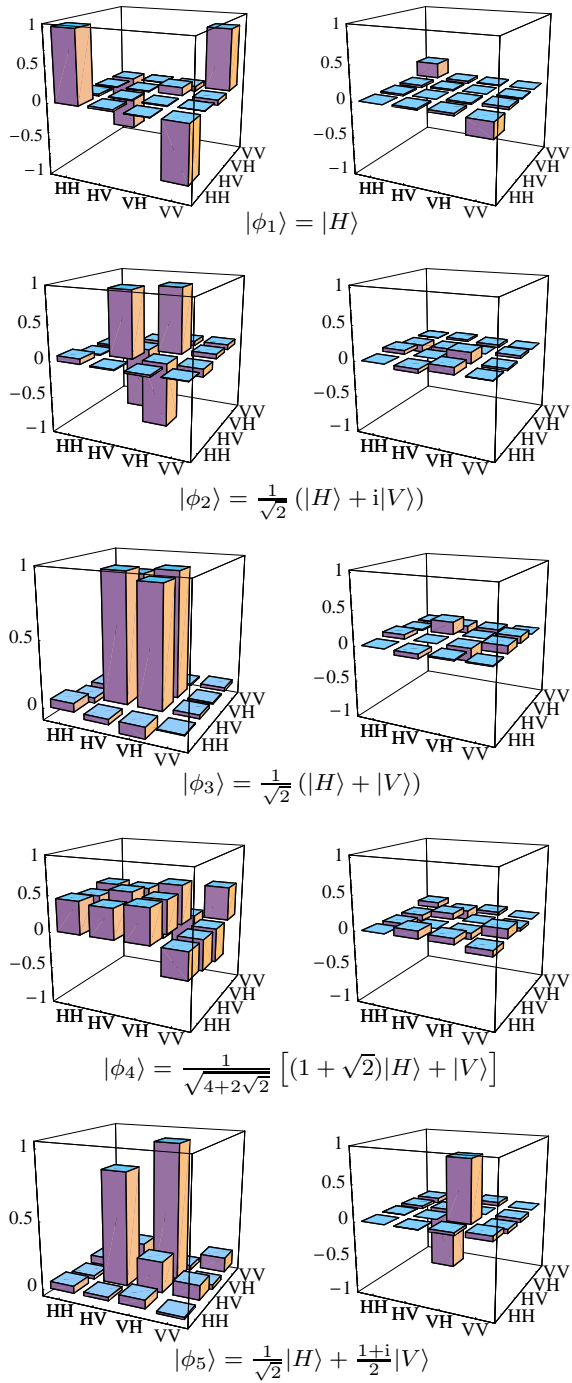}}}
\caption{The complete characterization of the programmable
SIGN gate operation by means of full quantum process tomography. The real
(left column) and imaginary (right column) parts of the reconstructed
process matrix $\chi$ are shown for five different program-register
states (\ref{program_register}) encoding the sign flip operation in bases
$\{|\phi_j\rangle$,\,$|{\phi_j}_{\perp}\rangle\}$, $j=1,\ldots,5$.}
\label{sigmagate_fig3}
\end{figure}

The input-output measurement and the full quantum process tomography was
carried out for five different entangled states of the program register
(\ref{program_register}) specified by the following single-qubit basis
states $|\phi_j\rangle$, $j=1,\ldots,5$, 
\begin{eqnarray}
& |\phi_1\rangle = |H\rangle, & \nonumber\\
& |\phi_2\rangle = |R\rangle = \frac{1}{\sqrt{2}}\left( |H\rangle + {\rm i}|V\rangle \right), & \nonumber\\
& |\phi_3\rangle = |D\rangle
  = \frac{1}{\sqrt{2}}\left( |H\rangle + |V\rangle \right), & \label{programstates}\\
& |\phi_4\rangle
  = \frac{1}{\sqrt{4+2\sqrt{2}}} \left[ (1+\sqrt{2}) |H\rangle + |V\rangle \right], & \nonumber\\
& |\phi_5\rangle = \frac{1}{\sqrt{2}} |H\rangle + \frac{1+{\rm i}}{2} |V\rangle. & \nonumber 
\end{eqnarray}
We quantify the performance of the implemented quantum gates by the process 
fidelity defined as follows,
\begin{equation}
  F(\chi,\chi_{\phi}) = \frac{\mathrm{Tr}[\chi \chi_{\phi}]}
 {\mathrm{Tr}[\chi]\mathrm{Tr}[\chi_{\phi}]}.
  \label{Fidelitychi}
\end{equation}
Here $\chi_{\phi}$ is a process matrix representing the
target SIGN gate $\Sigma_\phi$ for a  particular basis
$\{|\phi\rangle,|\phi_\perp\rangle\}$,
\begin{equation}
  \chi_{\phi} = \openone \otimes \Sigma_\phi |\Phi^{+}\rangle \langle \Phi^{+}|
  \openone \otimes \Sigma_\phi^{\dagger}.
\end{equation}
Note that $\chi_{\phi}$ is effectively a density matrix of a 
pure maximally entangled two-qubit state.

A detailed analysis of the reconstructed process matrices $\chi_j$, $j=1,\ldots,5,$
reveals that they exhibit an identical small unitary deviation $\delta \Sigma$  
from the programmed operation  $\Sigma_\phi$ due to insufficiently
compensated polarization transformation induced by optical fibers employed in BSM.
This deviation can be corrected by applying the inverse unitary operation
$\delta\Sigma^\dagger$ on the input photon $1$ using polarization
controller or numerically applying the fixed correcting unitary operation
to all the reconstructed process matrices. The optimal correcting
operation $\delta \Sigma^\dagger$ can be determined by maximizing
the average process fidelity $\bar{F}=\frac{1}{5}\sum_j
F(\tilde{\chi_j},\chi_{\phi_j})$, where 
\begin{equation}  \label{fixed_correction}
  \tilde{\chi}_j = \left( \delta\Sigma \otimes \openone \right) \chi_j
  \left( \delta \Sigma^\dagger \otimes \openone \right).
\end{equation}
We parameterize the $SU(2)$ matrix $\delta\Sigma$ by the three Euler angles and
numerically determine their optimal values which maximize the average process fidelity
$\bar{F}$. We emphasize that since the unitary operation $\delta\Sigma$ is fixed, 
it does not influence the programmability of the gate in any way.

\begin{table}[!t!]
\begin{ruledtabular}
\begin{tabular}{cccccc}
    basis & $|\phi_1\rangle$ & $|\phi_2\rangle$ & $|\phi_3\rangle$ &
    $|\phi_4\rangle$ & $|\phi_5\rangle$ \\
  \hline
     $F$        & 0.866 & 0.903 & 0.931 & 0.905 & 0.889 \\ % 0.899+-0.024
     ${\cal P}$ & 0.788 & 0.844 & 0.902 & 0.852 & 0.837 \\ % 0.85+-0.04
     $E_f$      & 0.712 & 0.798 & 0.889 & 0.816 & 0.780     % 0.80+-0.06
\end{tabular}
\end{ruledtabular}
\caption{Parameters of the implemented programmable SIGN gate
determined from the reconstructed process matrices shown in
Fig.~\ref{sigmagate_fig3} and corresponding to five different programs
specified by basis states $|\phi_j\rangle$. Shown are the quantum
process fidelity $F$ that attains the average value $0.90\pm0.03$,
the process purity ${\cal P}$ and the entanglement of formation
$E_f$ whose average values read $0.85\pm0.04$ and $0.80\pm0.06$.}
% , respectively.
\label{processfidelity_tab1}
\end{table}

The final reconstructed quantum processes corrected for the fixed unitary
offset $\delta\Sigma$ are shown in Fig.~\ref{sigmagate_fig3}. The process
fidelity determined from the reconstructed process matrices is shown
in Table~\ref{processfidelity_tab1}. The average process fidelity is
$0.90\pm0.03$ which demonstrates very good functionality and performance
of the programmable SIGN gate. Without the correction for $\delta\Sigma$,
the average fidelity of $0.85\pm0.03$ is reached. Another important
characterization of the implemented gates is provided by the purity
of the effective two-qubit state
$\chi$, $\mathcal{P}=\mathrm{Tr}[\chi^2]/(\mathrm{Tr}[\chi])^2$.
The purity quantifies how close is the transformation to a purity-preserving
operation  $\mathcal{E}(\rho)=E\rho E^\dagger$. The resulting values of
$\mathcal{P}$ given in Table~\ref{processfidelity_tab1} exceed $0.79$ and
this further confirms that the operations applied by the 
programmable gate are close to purity-preserving unitary transformations.
Finally, we also calculate the entanglement of formation $E_f$ 
of the process matrix $\chi$. This quantity characterizes the ability 
of the programmable gate to preserve entanglement. In particular, when the
gate is applied to one photon from a maximally entangled pair then $E_f$
will be the resulting entanglement of that photon pair. For all the
reconstructed process matrices it holds that $E_f>0.71$, c.f. Table~I,
so the programmable gate is capable to preserve a large fraction
of entanglement when applied to one part of an entangled state.

In summary, we have experimentally implemented a programmable
single-qubit quantum gate which can perform a rotation of a Bloch
sphere about $\pi$ radians about an arbitrary axis specified by
the quantum state of a program register. We have performed
a complete tomographic characterization of the gate for several
different programs and thus confirmed its programmability and
very good performance. To the best of our knowledge, this is the
first demonstration of a programmable quantum gate with a two-qubit
register, that is necessary for the considered class of encoded
operations. Our scheme can be straightforwardly generalized to fully
universal programmable gate that can apply arbitrary unitary operation 
on the data qubit and it paves the way towards realization of 
even more complex programmable quantum gates with multiphoton
programs. On more fundamental basis, our experiment confirms the
striking capability of few-qubit quantum states to contain complete
information on a quantum operation whose classical characterization
would require infinitely many classical bits. These unique features
of quantum states can be explored in teleportation-based quantum
computation \cite{Gottesman99} and our work can be seen as an
important enabling step in this direction.  

\begin{acknowledgments}
This work has been supported by the research projects No.~LC06007,
No.~MSM6198959213, and No.~1M06002 of the Czech Ministry of Education,
and No.~202/08/0224 of Czech Grant Agency.
\end{acknowledgments}

\end{document}